\documentstyle[12pt]{article}
\begin{document}

%%%%%%%%%%%%%%%%%%%%%%%%%%%%%%%%%%%%%%%%%%%%%%%%%%%%%%%%%%%%%%%%%%%%%%%%%%%%
%

 \renewcommand{\thesection}{\arabic{section}}
 \renewcommand{\thesubsection}{\thesection.\arabic{subsection}}
 \renewcommand{\theequation}{\arabic{equation}}
 \renewcommand {\c}  {\'{c}}
 \newcommand {\cc} {\v{c}}
 \newcommand {\s}  {\v{s}}
 \newcommand {\CC} {\v{C}}
 \newcommand {\C}  {\'{C}}
 \newcommand {\Z}  {\v{Z}}
 \baselineskip=24pt

 %%%%%%%%%%%%%%%%%%%%%%%%%%%%%%TITLE
 %%%%PAGE%%%%%%%%%%%%%%%%%%%%%%%%%%%%%%%%%%%%%%

 \begin{center}
 {\bf \Large New realizations of Lie algebra kappa-deformed\\
  Euclidean space}

 \bigskip

 Stjepan Meljanac\footnote{e-mail: meljanac@thphys.irb.hr}
 and Marko Stoji\'c \footnote{e-mail: marko.stojic@zg.htnet.hr}\\

 Rudjer Bo\v{s}kovi\'c Institute, Bijeni\v cka  c. 54, 10002
 Zagreb,
 Croatia

 \bigskip

 Short Title: New realizations of kappa-deformed Euclidean space

 \end{center}
 \setcounter{page}{1}
 \bigskip
 %%%%%%%%%%%%%%%%%%%%%%%%%%%%%%%%%%%%%%%%%%%%%%%%%%%%%%%%%%%%

 %%%%%%%%%%%%%%%%%%%%ABSTRACT
 %%%%%%%%%%%%%%%%%%%%%%%%%%%%%%%%%%%%%%%%%%%%%%%%%%%%%%%%%%%%

 {\bf Abstract.}

 We study Lie algebra  $\kappa$-deformed Euclidean space  with
 undeformed rotation algebra $SO_a(n)$ and commuting vectorlike
 derivatives. Infinitely many realizations in terms of commuting coordinates
 are constructed and a corresponding star product is found for
 each of them. The $\kappa$-deformed noncommutative  space of the Lie algebra
 type with undeformed Poincar{\'e} algebra and with the corresponding
 deformed coalgebra is constructed in a unified way.

 \newpage

%%%%%%%%%%%%%%%%%%%%%%%%%%%%%%%%%%%%%%%%%%%%%%%%%%%%%%%%%%%%%%%%%%%%%%%%%%%%

%%%%%%%%%%%%%%%%%%%%%%%%%%%% SECTION 1:

%%%%%%%%%%%%%%%%%%%%%%%%%%%%%%%%%%%%%%%%%%%%%%%%%%%%%%%%%%%%%%%%%%%%%%%%%%%%

 \section {Introduction}

 %%%%%%%%%%%%%%%%%%%%%%%%%%%%%%%%%%%%%%%%%%%%%%%%%%%%%%%%%%%%%%%%%%%%%%%%
 %%%%%%%%%%%%%%%%%%%%%%%%%%%%%%%%%%%%%%%%%%%%%%%%%%%%%%%%%%%%%%%%%%%%%

 In the last decade, there has been a great interest
 in the formulation and consistency  of physical theories
 defined on noncommutative (NC) spaces, and in finding their consequences
 [1-5].\\
 However,  there is no clear guiding physical principle how to build
 a fundamental theory on NC spaces and which NC spaces are physically
 acceptable and preferable.
 Also, it is not known  how matter and gravity influence the properties of
 NC spaces at small distances, and vice versa.

 Nevertheless, it is important to classify NC spaces and their
 properties, and, particularly, to develop a unifying approach to and
 a generalized theory  for such NC spaces convenient for physical applications.
 The notion of generalized symmetries and their role in the analysis of NC spaces
 is also crucial.
 In order to make a step in this direction, we analyze a NC space of the
 Lie algebra type, particularly the so-called
 $\kappa$-deformed space introduced in [6-8].

 For simplicity, we  restrict ourselves  to $\kappa$-deformed
 Euclidean space. The analysis can be easily extended to
 $\kappa$-deformed Minkowski space.
 The dimensional parameter $a=\frac{1}{\kappa}$ is a very small length
 scale, and when it goes to zero, the  undeformed space appears
 as a smooth limit.
 The generators of generalized rotations satisfy undeformed $
 SO_a(n)$ algebra, i.e. undeformed Lorentz algebra in
 $\kappa$-deformed Minkowski space.
 Dirac derivatives are assumed to mutually commute and transform as a vector
 representation under $SO_a(n)$ algebra.
 This $\kappa$-deformed space was studied by different groups, from both
 mathematical and physical point of view [9-18].
 Specially, realizations in terms of commutative coordinates  were
 obtained  and  discussed  in the cases of symmetric  ordering  and normal
 (left and right) ordering of NC coordinates [12, 18].

 We analyze  $\kappa$-deformed Euclidean space using the methods
 developed for deformed single and multimode oscillators in the Fock space
 representations [19-26].
 Particularly, we use the methods for constructing deformed
 creation and annihilation  operators in terms of ordinary bosonic
 multimode oscillators, i.e. a kind of bosonisation  [19,20,25].
 Also, we use the construction  of transition number operators and,
 generally, of generators proposed in Refs.~[20,21,24].\\
 The simple connection between creation and
 annihilation operators with NC coordinates and Dirac derivatives is
 established by the Bargmann-type representation. We find infinitely
 many new realizations in terms of commutative coordinates.\\
 All these realizations are on an equal footing, and  a star
 product is associated to each of them.
 The general feature of NC spaces is that there are generally
 infinitely many realizations in terms of commutative coordinates
 and the physical results should not depend on the realization used.

 The plan of the paper is as follows. In section 2 we present
 $\kappa$-deformed Euclidean space and its realizations in  Euclidean space.
 In section 3 the undeformed rotation algebra $SO_a(n)$
 compatible with kappa deformation and its general realizations
 are considered.
 In  section 4 the action of the generators $SO_a(n)$ on NC
 coordinates leads to infinitely many new realizations of NC space
 in terms of commutative coordinates and their derivatives.
 In  section 5 the corresponding realizations of Dirac derivatives
 are constructed. The Lie algebra type $\kappa$-deformed NC space with
 undeformed rotation (Poincar{\'e}) algebra and the
 corresponding deformed coalgebra is proposed in a unique way.
 In  section 6 the invariant Klein-Gordon operator and its
 realizations are given for  $\kappa$-deformed Euclidean space,
 with a short summary of all realizations included.
 In section 7 the hermiticity properties are discussed. In
 section 8 the corresponding realizations for star products are
 presented.
 Finally, in  section 9  a short conclusion is given.

%%%%%%%%%%%%%%%%%%%%%%%%%%%%%%%%%%%%%%%%%%%%%%%%%%%%%%%%%%%%%%%%%%%%%%%%%%%%
%%%%%%%%%%%%%%%%%%%%%%%%%%%SECTION 2:
%%%%%%%%%%%%%%%%%%%%%%%%%%%%%%%%%%%%%%%%%%%%%%%%%%%%%%%%%%%%%%%%%%%%%%%%%%%%

\section { Kappa-deformed Euclidean space \\ and its realizations}

%%%%%%%%%%%%%%%%%%%%%%%%%%%%%%%%%%%%%%%%%%%%%%%%%%%%%%%%%%%%%%%%%%%%%%%%%
%%%%%%%%%%%%%%%%%%%%%%%%%%%%%%%%%%%%%%%%%%%%%%%%%%%%%%%%%%%%%%%%%%%%%%%%%%%

Let us consider a Lie algebra type  noncommutative (NC) space with
coordinates $\hat{x}_{1},\hat{x}_{2},...,\hat{x}_{n},$ as
\begin{equation}
\lbrack\hat{x}_{\mu},\hat{x}_{\nu}\rbrack=
iC_{\mu\nu\lambda}\hat{x}_{\lambda} =
i(a_{\mu}\hat{x}_{\nu}-a_{\nu}\hat{x}_{\mu}),
\end{equation}
where $ a_{1},a_{2},...,a_{n}$ are  constant real parameters
describing a deformation of Euclidean space. The structure
constants are
\begin{equation}
C_{\mu\nu\lambda}=a_{\mu}\delta_{\nu\lambda}-a_{\nu}\delta_{\mu\lambda}.
\end{equation}
We choose   $ a_{1}=a_{2}=...=a_{n-1}=0, a_{n}=a $ and use  Latin
indices  for the subspace $(1,2,...,n-1)$ and Greek indices  for
the whole space $(1,2,...,n)$, see [16-18]. Then the algebra of
the NC coordinates becomes
\begin{equation}
\lbrack\hat{x}_{i},\hat{x}_{j}\rbrack=0, \qquad
\lbrack\hat{x}_{n},\hat{x}_{i}\rbrack=ia\hat{x}_{i},\qquad
i,j=1,2,...,n-1.
\end{equation}
Using the methods developed in [19,20,24,25] and the Bargmann
representation we point out that there exists a realization of the
NC coordinates $\hat{x}_{\mu}$ in terms of ordinary commutative
coordinates  $ x_{1},x_{2},...,x_{n}$ and their derivatives
$\partial_{1},\partial_{2},...,\partial_{n},$  where
$\partial_{\mu}=\frac{\partial}{\partial x_{\mu}}.$ The most
general Ansatz for the NC coordinates $\hat{x}_{\mu}$ satisfying
the algebra (3)  is
\begin{equation}
\begin{array}{c}
\hat{x}_{i}=x_{i}\varphi(A),\\
\\
\hat{x}_{n}=x_{n}\psi(A)+iax_{k}\partial_{k}\gamma(A),\\
\end{array}
\end{equation}
where  $A=ia\partial_{n}$. In the above relations the deformed
creation operators are represented by $\hat{x}_\mu$, the bosonic
creation operators by $ x_{\mu}$, the bosonic annihilation
operator by  $\partial_{\mu}$ and the vacuum state by $1$.
Inserting this Ansatz into  eq. (3) we obtain
\begin{equation}
\frac{\varphi'}{\varphi}\psi=\gamma-1,
\end{equation}
where $\varphi'=\frac{d\varphi}{dA}$.\\
There are infinitely many representations parametrized by two of
the functions $ \varphi,\psi, \gamma $, with the boundary
conditions
\begin{equation}
\varphi(0)=1,\quad  \psi(0)=1,
\end{equation}
 and with $\gamma(0)=\varphi'(0)+1$  finite.
At this point we could make  rather ad hoc arbitrary assumptions
on the  derivatives $\hat\partial_{\mu}$ which can not be
motivated neither physically nor mathematically.

%%%%%%%%%%%%%%%%%%%%%%%%%%%%%%%%%%%%%%%%%%%%%%%%%%%%%%%%%%%%%%%%%%%%%
%%%%%%%%%%%%%%%%%%%%%%%%%%%%%SECTION 3:
%%%%%%%%%%%%%%%%%%%%%%%%%%%%%%%%%%%%%%%%%%%%%%%%%%%%%%%%%%%%%%%%%%%%

\section  {$SO_a(n)$  algebra}
%%%%%%%%%%%%%%%%%%%%%%%%%%%%%%%%%%%%%%%%%%%%%%%%%%%%%%%%%%%%%%%%%%%%
%%%%%%%%%%%%%%%%%%%%%%%%%%%%%%%%%%%%%%%%%%%%%%%%%%%%%%%%%%%%%%%%%%%%

Instead, we demand that there should exist generators $M_{\mu\nu}$
satisfying the ordinary undeformed $SO_a(n)$ algebra:
\begin{equation}
\lbrack M_{\mu\nu},M_{\lambda\rho}\rbrack =
\delta_{\nu\lambda}M_{\mu\rho}-\delta_{\mu\lambda}M_{\nu\rho}
-\delta_{\nu\rho}M_{\mu\lambda}+\delta_{\mu\rho}M_{\nu\lambda}.
\end{equation}

There are infinitely many representations of  $M_{\mu\nu}$ in
terms of commutative  coordinates $x_{\lambda},$ their derivatives
$\partial_{\lambda}$, and deformation parameters
$a_{1},a_{2},...,a_{n}$. The generators  $M_{\mu\nu}$ are linear
in $x$ and form an infinite series in $\partial$.\\
Let us assume that $a_i=0$, $a_{n}=a,$ i.e. the same deformation
parameters as in  the $\hat{x}$  NC coordinate algebra, eq. (3).
Then a simple Ansatz is
\begin{equation}
\begin{array}{c}
M_{ij}=x_{i}\partial_{j}-x_{j}\partial_{i},\\
\\
M_{in}=x_{i}\partial_{n}F_{1}-x_{n}\partial_{i}F_{2}+ia
x_{i}\Delta F_{3}+ia x_{k}\partial_{k}\partial_{i}F_{4},\\
\\
M_{ni}=-M_{in},
\end{array}
\end{equation}
where $\Delta=\partial_{k}\partial_{k}$ and the summation over
repeated indices is understood. The functions
$F_{1},F_{2},F_{3},F_{4} $ depend on $A=ia\partial_{n}$. In
principle, the $F$  functions could depend on $B=(ia)^2\Delta$.
For simplicity, we assume that $F$ dependen on $A$ only.\\
Inserting  Ansatz~(8), into the algebra (7), from
\begin{equation}
\begin{array}{c}
\lbrack M_{ij},M_{jn}\rbrack=M_{in},\\
\\
\lbrack M_{in},M_{jn}\rbrack=-M_{ij},
\end{array}
\end{equation}
we obtain the following two equations:
\begin{equation}
\begin{array}{c}
F_{1}F_{2}+A F_{1}'F_{2}+
AF_{1}F_{4}-2AF_{1}F_{3}=1,\\
\\
2F_{3}^2-F_{3}'F_{2}+F_{3}F_{4}=0,
\end{array}
\end{equation}
where $F'=\frac{dF}{dA}$. Note that two of  the functions
$F_{1},F_{2},F_{3},F_{4}$ are arbitrary. The boundary conditions
are
\begin{equation}
F_{1}(0)=F_{2}(0)=1,
\end{equation}
and $F_{3}(0)$,  $F_{4}(0)$ are required to be finite.

Now we can calculate the commutators $\lbrack
M_{\mu\nu},\hat{x}_{\lambda}\rbrack$, substituting
$\hat{x}_\lambda$, eq. (4), and $M_{\mu\nu}$,  eq. (8). The result
is expressed in terms of $\varphi$, $\psi$, $\gamma$ and
$F$-functions, restricted by eqs.~(5), (10), and is linear in the
commutative coordinates $x_\mu$. In general, this result can not
be expressed in terms of $\hat{x}$ and $M$ only, without the
derivatives $\partial$.
%%%%%%%%%%%%%%%%%%%%%%%%%%%%%%%%%%%%%%%%%%%%%%%%%%%%%
%%%%%%%%%%%%%%%%%%%%%%%%%%%%%%SECTION 4:
%%%%%%%%%%%%%%%%%%%%%%%%%%%%%%%%%%%%%%%%%%%%%%%%%%%%%%%

\section {Action of  SO$_a$(n) generators on NC coordinates}
%%%%%%%%%%%%%%%%%%%%%%%%%%%%%%%%%%%%%%%%%%%%%%%%%%%%%%%%%%
%%%%%%%%%%%%%%%%%%%%%%%%%%%%%%%%%%%%%%%%%%%%%%%%%%%%%%%%%%%%%%%

At this point we  demand that the generators $M_{\mu\nu}$,
$\hat{x}_{\lambda}$  close linearly under  commutation, i.e. we
obtain the extended Lie algebra with extended structure constants
satisfying Jacobi relations. In order to construct the extended
Lie algebra of generators $\hat{x}_{\lambda}$, and $M_{\mu\nu}$,
for general $a_{\mu}$, we proceed as follows. The most general
covariant form of commutators $ \lbrack
M_{\mu\nu},\hat{x}_{\lambda}\rbrack,$ of the generators of
rotations $M_{\mu\nu}$ with NC coordinates $\hat{x}_{\lambda},$ is
antisymmetric in indices $\mu$ and $\nu$, linear in generators
$\hat{x}$, $M$, and with smooth limit
$[M_{\mu\nu},x_{\lambda}]=x_{\mu}\delta_{\nu\lambda}-x_{\nu}\delta_{\mu\lambda},$
when $a_{\mu}\rightarrow 0.$  It is given by
$$
\lbrack M_{\mu\nu},\hat{x}_{\lambda}\rbrack =
\hat{x}_{\mu}\delta_{\nu\lambda}-\hat{x}_{\nu}\delta_{\mu\lambda}
+isa_{\lambda}M_{\mu\nu}-it(a_{\mu}M_{\nu\lambda}-a_{\nu}M_{\mu\lambda})
+iua_\alpha (M_{\alpha\mu}\delta_{\nu\lambda}-
M_{\alpha\nu}\delta_{\mu\lambda}) ,
$$
where $s,t,u\in\mathbf{R}$.\\ The necessary and sufficient
conditions for the consistency of extended Lie algebra with
generators $\hat{x}_{\lambda}$ and $M_{\mu\nu}$ are:
$$
\lbrack M_{\alpha\beta},\lbrack
\hat{x}_{\mu},\hat{x}_{\nu}\rbrack\rbrack+ \lbrack
\hat{x}_{\mu},\lbrack \hat{x}_{\nu},M_{\alpha\beta}\rbrack\rbrack+
\lbrack\hat{x}_{\nu},\lbrack
M_{\alpha\beta},\hat{x}_{\mu}\rbrack\rbrack=0,
$$
$$
\lbrack M_{\alpha\beta},\lbrack
M_{\gamma\delta},\hat{x}_{\mu}\rbrack\rbrack+ \lbrack
M_{\gamma\delta},\lbrack
\hat{x}_{\mu},M_{\alpha\beta}\rbrack\rbrack+
\lbrack\hat{x}_{\mu},\lbrack
M_{\alpha\beta},M_{\gamma\delta}\rbrack\rbrack=0.
$$
All extended Jacobi identities are satisfied for the unique
solution $s=u=0$, $t=1$:
\begin{equation}
\lbrack M_{\mu\nu},\hat{x}_{\lambda}\rbrack =
\hat{x}_{\mu}\delta_{\nu\lambda}-\hat{x}_{\nu}\delta_{\mu\lambda}
-ia_{\mu}M_{\nu\lambda}+ia_{\nu}M_{\mu\lambda}.
\end{equation}
Inserting $a_{i}=0 $,  $ a_{n}=a $ we obtain two important
relations:
\begin{equation}
\lbrack M_{in},\hat{x}_{n}\rbrack=\hat{x}_{i}+ia M_{in},
\end{equation}
\begin{equation}
\lbrack
M_{in},\hat{x}_{j}\rbrack=-\delta_{ij}\hat{x}_{n}+iaM_{ij}.
\end{equation}
An important ingredient of the symmetry structure of
$\kappa$-deformed  space are the Leibniz  rules of generators of
rotations. They can be derived immediately from eqs. (13) and
(14), (see [16-18])
$$
M_{ij}(f\cdot g)=(M_{ij}f)\cdot g+f\cdot(M_{ij}g),\\
$$
$$
M_{in}(f\cdot g)=(M_{in}f)\cdot
g+(e^Af)\cdot(M_{in}g)+ia(\partial_j\frac{1}{\varphi(A)}f)\cdot(M_{ij}g),
$$
where $f$, $g$ are functions of  NC coordinates $\hat{x}_{\mu}$,
and
$$
\lbrack\partial_i,\hat{x}_j\rbrack=\delta_{ij}\varphi(A), \qquad
\lbrack\partial_i,\hat{x}_n\rbrack=ia\partial_i\gamma(A),
$$
$$
\lbrack\partial_n,\hat{x}_i\rbrack=0,\qquad
\lbrack\partial_n,\hat{x}_n\rbrack=1
$$
for $\psi=1$.\\
In a more technical language, the above equations are the
coproducts
$$
\triangle M_{ij}=M_{ij}\otimes 1+1\otimes M_{ij},
$$
$$
\triangle M_{in}=M_{in}\otimes 1+e^A\otimes M_{in}+iaD_je^A\otimes
M_{ij},
$$
where $e^A$ and Dirac derivatives $D_{\mu}$ are defined in Section
5, eq. (29), for the case $\psi=1$. The final result for coproduct
depends only on $aD_{\mu}$, and $a^2D_{\mu}D_{\mu}$. In the limit
$a\rightarrow 0$  it gives ordinary undeformed coproduct for
$M_{\mu\nu}.$
%The notion of coproduct leads to the observation
%that the generators of the $\kappa$-deformed symmetry are elements
%of a
%Hopf algebra.\\
The coproduct $\Delta$, which we determined for $M_{\mu\nu}$,
multiplicatively extends to the whole algebra $SO_a(n)$, which
becomes a Hopf algebra in this way.

From the relation (13) we obtain four equations:
\begin{equation}
\begin{array}{c}
F_{1}\psi+AF_{1}'\psi-AF_{1}(\gamma+1)-\varphi=0,\\
\\
F_{2}'\psi-F_{2}\psi'+F_{2}(\gamma-1)=0 , \\
\\
F_{3}'\psi+F_{3}(\gamma-1)=0 , \\
\\
F_{4}'\psi+F_{2}\gamma'+F_{4}(\gamma-1)=0,  \\
\end{array}
\end{equation}
and from the relation (14) we obtain two equations:
\begin{equation}
\begin{array}{c}
F_{2}\varphi'+F_{4}\varphi+1=0,  \\
\\
F_{3}=\frac{1}{2\varphi}.
\end{array}
\end{equation}
Now we have six additional equations, i.e. eight equations (10),
(15), (16) for four  functions $F_1$, $F_2$, $F_3$ and $F_4$.
Hence, there are four additional equations, which have to be
satisfied. From these consistency relations we obtain two infinite
families of solutions
satisfying simultaneously (10), (15) and (16).\\
{\bf I realization: $\psi=1$}
\begin{equation}
\begin{array}{c}
F_{1}=\varphi\frac{e^{2A}-1}{2A} , \quad F_{2}=\frac{1}{\varphi} ,
\quad  F_{3}=\frac{1}{2\varphi}, \quad
F_{4}=-\frac{\gamma}{\varphi},
\end{array}
\end{equation}
where
\begin{equation}
\gamma=\frac{\varphi'}{\varphi}+1.
\end{equation}
{\bf II realization: $\psi=1+2A$}
\begin{equation}
\begin{array}{c}
F_{1}=\varphi , \quad F_{2}=\frac{\psi}{\varphi} , \quad
F_{3}=\frac{1}{2\varphi} , \quad F_{4}=-\frac{\gamma}{\varphi},
\end{array}
\end{equation}
where
\begin{equation}
\varphi=\frac{C-1}{C-\sqrt{\psi}} , \quad
\gamma=\psi\frac{\varphi'}{\varphi}+1, \quad
C\in{\mathbf{R}},\quad C\neq1.
\end{equation}
The first realization $\psi=1$ can be parametrized by an arbitrary
function $\varphi(A),$ \quad $ \varphi(0)=1.$ The second
realization $\psi=1+2A$ is parametrized with $C\in$ $\mathbf{R}$,
$C\neq1.$

%%%%%%%%%%%%%%%%%%%%%%%%%%%%%%%%%%%%%%%%%%%%%%%%%%%%%%%%%%%%%%%%%%%
%%%%%%%%%%%%%%%%%%%%%%%%%%%%%%%%SECTION 5:
%%%%%%%%%%%%%%%%%%%%%%%%%%%%%%%%%%%%%%%%%%%%%%%%%%%%%%%%%%%%%%%%%%

%%%%%%%%%%%%%%%%%%%%%%%%%%%%%%%%%%%%%%%%%%%%%%%%%%%%%%%%%%%%%%%
\section {Dirac derivatives}
%%%%%%%%%%%%%%%%%%%%%%%%%%%%%%%%%%%%%%%%%%%%%%%%%%%%%%%%%%%%
%%%%%%%%%%%%%%%%%%%%%%%%%%%%%%%%%%%%%%%%%%%%%%%%%%%%%%%%%%%%%%%

Imposing the undeformed $SO_a(n)$ algebra it is natural to define
the Dirac derivatives  $D_{\mu}$ as
\begin{equation}
\begin{array}{c}
\lbrack M_{\mu\nu},D_{\lambda}\rbrack
=\delta_{\nu\lambda}D_{\mu}-\delta_{\mu\lambda}D_{\nu},\\
\\
\lbrack D_{\mu},D_{\nu}\rbrack=0.\\
\end{array}
\end{equation}
The most general Ansatz corresponding to $a_{i}=0,$ $a_{n}=a$ is
\begin{equation}
\begin{array}{c}
D_{i}=\partial_{i}G_{1}(A),\\
\\
D_{n}=\partial_{n}G_{2}(A)+ia\Delta G_{3}(A).
\end{array}
\end{equation}
Inserting them into
\begin{equation}
\begin{array}{c}
\lbrack M_{in},D_{n}\rbrack=D_{i},\\
\\
\lbrack M_{in},D_{i}\rbrack=-D_{n},
\end{array}
\end{equation}
we find five equations:
\begin{equation}
\begin{array}{c}
AF_{2}G_{2}'+F_{2}G_{2}-2AF_{1}G_{3}-G_{1}=0,\\
\\
F_{2}G_{3}'-2(F_{3}+F_{4})G_{3}=0 , \\
\\
F_{1}G_{1}-G_{2}=0,\\
\\
F_{2}G_{1}'-F_{4}G_{1}=0.  \\
\\
F_{3}G_{1}-G_{3}=0.  \\
\end{array}
\end{equation}
The boundary conditions are
\begin{equation}
G_{1}(0)=1, \quad G_{2}(0)=1,
\end{equation}
and $G_{3}(0)$ is required to be finite.
Using our realizations for the F-functions, eqs. (17)-(20), we find \\
{\bf I realization: $\psi=1$ }
\begin{equation}
\begin{array}{c}
G_{1}=\frac{e^{-A}}{\varphi}, \quad G_{2}=\frac{sinhA}{A}, \quad
G_{3}=\frac{e^{-A}}{2\varphi^2}.
\end{array}
\end{equation}
{\bf II realization: $\psi=1+2A$}
\begin{equation}
\begin{array}{c}
G_{1}=\frac{C-\sqrt\psi}{(C-1)\sqrt\psi}, \quad
G_{2}=\frac{1}{\sqrt\psi}, \quad G_{3}=\frac{\lgroup C-\sqrt\psi
\rgroup ^2}{2(C-1)^2\sqrt\psi}.
\end{array}
\end{equation}
Now we calculate the commutation relations between NC coordinates
$\hat{x}_{\mu}$ and the Dirac derivatives $D_{\nu}$:
\begin{equation}
\begin{array}{c}
\lbrack
D_{i},\hat{x}_{j}\rbrack=\delta_{ij}(-iaD_{n}+\sqrt{1-a^2D_{\mu}D_{\mu}}),\\
\\

\lbrack D_{i},\hat{x}_{n}\rbrack=0 , \\
\\
\lbrack D_{n},\hat{x}_{i}\rbrack=iaD_{i},\\
\\
\lbrack D_{n},\hat{x}_{n}\rbrack=\sqrt{1-a^2D_{\mu}D_{\mu}}.
\end{array}
\end{equation}
These relations are universal for both realizations, $\psi=1$ and
$\psi=1+2A,$ and they involve only the deformation parameter
$a$.\\
The corresponding coproduct is given by (see [16-18]):
$$
\triangle
D_n=D_n\otimes\Big(-iaD_n+\sqrt{1-a^2D_{\mu}D_{\mu}}\Big)+
\frac{iaD_n+\sqrt{1-a^2D_{\mu}D_{\mu}}}{1-a^2D_k D_k}\otimes D_n
$$
$$
 +iaD_i\frac{iaD_n+\sqrt{1-a^2D_{\mu}D_{\mu}}}{1-a^2D_kD_k}\otimes D_i,
$$
$$
\triangle
 D_i=D_i\otimes\Big(-iaD_n+\sqrt{1-a^2D_{\mu}D_{\mu}}\Big)+1\otimes D_i.
$$
The Dirac derivatives, together with the generators of rotations
$M_{\mu\nu}$, form a $\kappa$-deformed Euclidean Hopf algebra
which is undeformed in the algebra sector, eqs. (7), (21). The
deformation is purely in the coalgebra sector following from eqs.
(13), (14) and (28). Namely,  the coalgebra structure is
determined from the deformed commutation relations, eqs. (13),
(14), (28), in a simple and unique way.

In the first realization $\psi=1$ one finds the relation
\begin{equation}
e^{-A}=-iaD_{n}+\sqrt{1-a^2D_{\mu}D_{\mu}}
\end{equation}
and in the second realization $\psi=1+2A $ the following relation
holds:
\begin{equation}
\frac{1}{\sqrt{1+2A}}=-iaD_{n}+\sqrt{1-a^2D_{\mu}D_{\mu}}.
\end{equation}
Note that the relations $[D_\mu,\hat{x}_\nu],$ eq. (28), and  the
vacuum condition $D_\mu|0>=0$ define the Fock space. The  Gramm
matrices in the Fock space [20,22] imply the relations
$[\hat{x}_\mu,\hat{x}_\nu],$ eq. (3), and $[D_\mu,D_\nu]=0$.\\
We point out that the $\kappa$-deformed NC space of the Lie
algebra type with the structure constants
$C_{\mu\nu\lambda}=a_\mu\delta_{\nu\lambda}-a_\nu\delta_{\mu\lambda}$,
(eqs. (1), (2))
\begin{equation}
\lbrack\hat{x}_{\mu},\hat{x}_{\nu}\rbrack=iC_{\mu\nu\lambda}\hat{x}_{\lambda},
\end{equation}
together with the undeformed $SO_a(n)$ rotation algebra, eq.~(7),
leads to the relations
\begin{equation}
\lbrack M_{\mu\nu},\hat{x}_{\lambda}\rbrack =
\hat{x}_{\mu}\delta_{\nu\lambda}-\hat{x}_{\nu}\delta_{\mu\lambda}
-iC_{\mu\nu\alpha}M_{\alpha\lambda}.
\end{equation}
Demanding that the Dirac derivatives commute pairwise and
transform under a vector representation of $SO_a(n)$, eq.~(21), we
obtain a universal commutation relation
\begin{equation}
\lbrack
D_{\mu},\hat{x}_{\nu}\rbrack=\delta_{\mu\nu}\sqrt{1-a^2D_{\alpha}D_{\alpha}}+
iC_{\mu\alpha\nu}D_\alpha.
\end{equation}
Eqs. (12), (28) are unified in the above eqs. (32), (33). The Fock
space representation is defined by
$D_\mu|0>=0$,\quad $\forall\mu$.\\
The above NC space of the Lie algebra type with undeformed
$SO_a(n)$ algebra, and commuting  Dirac derivatives, transforming
as a vector representation under $SO_a(n)$, and with a smooth
limit to Euclidean space, is unique. The realizations of NC space
defined by  eqs. (7), (21), (31), (32), (33), and their
properties, will be treated separately. Some examples of the
Poincar{\'e} invariant interpretation of NC spaces and twisted
Poincar{\'e} coalgebra were also considered in [27-29].\\
Furthermore, it is interesting that, for $n=1$, the relation
$[D_n,\hat{x}_n]$ becomes
$$
[D,\hat{x}]=\sqrt{1-a^2D^2}.
$$
In the quadratic approximation in $a$, $[D,\hat{x}]\approx
1-\frac{1}{2}a^2D^2$. This commutation relation leads to the
generalized uncertainty relation with minimal length
$\frac{|a|}{\sqrt{2}}$ (see [30]).

%%%%%%%%%%%%%%%%%%%%%%%%%%%%%%%%%%%%%%%%%%%%%%%%%%%%%%%%%%
%%%%%%%%%%%%%%%%%%%%%%%%%%%%%SECTION 6:
%%%%%%%%%%%%%%%%%%%%%%%%%%%%%%%%%%%%%%%%%%%%%%%%%%%%%%

\section {Invariant operators}
%%%%%%%%%%%%%%%%%%%%%%%%%%%%%%%%%%%%%%%%%%%%%%%%%%%%%%%%%%%%%%%%%%
%%%%%%%%%%%%%%%%%%%%%%%%%%%%%%%%%%%%%%%%%%%%%%%%%%%%%%%%%%%%%%%%%

Analogously as we have defined Dirac derivatives, we introduce an
invariant operator $\Box$, generalizing the Laplace (D'Alambert)
operator, by the equation
\begin{equation}
\lbrack M_{\mu\nu},\Box\rbrack=0.
\end{equation}
A simple Ansatz is
\begin{equation}
\Box=\Delta H_{1}(A)+\partial_{n}^2H_{2}(A),
\end{equation}
with boundary conditions
$$
 H_{1}(0)=1, \quad H_{2}(0)=1.
$$
Then, from the relation (34), we obtain two equations:\\
\begin{equation}
\begin{array}{c}
AF_{2}H_{2}'+2F_{2}H_{2}-2F_{1}H_{1}=0,\\
\\
F_{2}H_{1}'-2(F_{3}+F_{4})H_{1}=0.
\end{array}
\end{equation}
\\
The solutions for the first realization  $\psi=1$ are
\begin{equation}
H_{1}=\frac{e^{-A}}{\varphi^2}, \quad H_{2}=-\frac{2\lbrack
1-coshA\rbrack}{A^2}.
\end{equation}
The second realization is for $\psi=1+2A:$
\begin{equation}
H_{1}=\frac{1}{\sqrt\psi\varphi^2}, \quad
H_{2}=\frac{2}{A^2}(\frac{1+A}{\sqrt{1+2A}}-1).
\end{equation}
Alternatively, we define another invariant operator
\begin{equation}
D_{\mu}D_{\mu}=D_{i}D_{i}+D_{n}D_{n}.
\end{equation}
For both realizations, $\psi=1$ and $\psi=1+2A$, the universal
relations
\begin{equation}
\begin{array}{l}
D_{\mu}D_{\mu}=\Box(1-\frac{a^2}{4}\Box),\\
\left[\Box,\hat{x}_\mu\right]=2D_\mu,
\end{array}
\end{equation}
hold, depending only on the deformation parameter $a$. The
corresponding Leibniz rule for $\Box(f\cdot g)$  can easily be
derived, Ref.~[18]. Note that there are infinitely many Dirac
derivatives and Laplace operators, differing by
multiplication by the function $ \phi(a^2\Box)$ with $\phi(0)=1$.\\
{\bf The new realizations can be summarized as follows:}\\
{\bf I realization: $\psi=1$}
\begin{equation}
\begin{array}{c}
M_{in}=x_{i}\partial_{n}\varphi\frac{e^{2A}-1}{2A}-x_{n}\partial_{i}\frac{1}{\varphi}+ia
x_{i}\Delta\frac{1}{2\varphi}-ia
x_{k}\partial_{k}\partial_{i}\frac{\gamma}{\varphi},\\
\\
D_{i}=\partial_{i}\frac{e^{-A}}{\varphi}, \qquad
D_{n}=\partial_{n}\frac{sinhA}{A}+ia\Delta\frac{e^{-A}}{2\varphi^2},\\
\\
\Box=\Delta\frac{e^{-A}}{\varphi^2}-\partial_{n}^2\frac{2\lbrack
1-coshA\rbrack}{A^2},\\
\end{array}
\end{equation}
where
$$
\gamma=\frac{\varphi'}{\varphi}+1.
$$
%%%%%%%%%%%%%%%%%%%%%%%%%%%%%%%%%%%%%%%%%%%%%%%%%
{\bf II realization: $\psi=1+2A$}

%%%%%%%%%%%%%%%%%%%%%%%%%%%%%%%%%%%%%%%%%%%%%%%%%%%%
\begin{equation}
\begin{array}{c}
M_{in}=x_{i}\partial_{n}\varphi-x_{n}\partial_{i}\frac{\psi}{\varphi}+ia
x_{i}\Delta\frac{1}{2\varphi}-ia
x_{k}\partial_{k}\partial_{i}\frac{\gamma}{\varphi},\\
\\
D_{i}=\partial_{i}\frac{C-\sqrt\psi}{(C-1)\sqrt\psi}, \qquad
D_{n}=\partial_{n}\frac{1}{\sqrt\psi}+ia\Delta\frac{\lgroup
C-\sqrt\psi\rgroup ^2}{2(C-1)^2\sqrt\psi},\\
\\
\Box=\Delta\frac{1}{\sqrt\psi\varphi^2}+
\partial_{n}^2\frac{2}{A^2}(\frac{1+A}{\sqrt{1+2A}}-1),
\end{array}
\end{equation}
where
$$
\varphi=\frac{C-1}{C-\sqrt{\psi}} , \quad
\gamma=\psi\frac{\varphi'}{\varphi}+1, \quad
C\in{\mathbf{R}},\quad C\neq 1.
$$

We point out, that starting from the above two realizations we can
construct infinitely many new realizations by similarity
transformations (i.e., composing our realizations with inner
automorphisms of the completed Weyl algebra of $x$-s,$\partial$-s)
$$
(\hat{x}_{\mu})_S=S\hat{x}_{\mu}S^{-1}, \qquad
(M_{\mu\nu})_S=SM_{\mu\nu}S^{-1},
$$
where
$$
S=exp\{\Phi(a\partial_1,...,a\partial_n)\}
$$
 with
 $$
\Phi(a\partial_1,...,a\partial_n) =
\sum_{\{m\}}c_{\{m\}}(x,\partial)\prod_{\mu=1}^n(a\partial_{\mu})^{m_\mu}
$$
and
$$
[x_{\mu}\partial_{\mu}, c_{\{m\}}(x,\partial)]=0,
$$
where $\{m\}=(m_1, m_2,..., m_n)$. To preserve smooth limit
$(\hat{x}_\mu)_S\rightarrow x_{\mu}$ when $a \rightarrow 0$, the
boundary condition on $\Phi(a\partial_1,...,a\partial_n)$  has to
be   $\Phi(0,...,0)=0,$  i.e.  $S\rightarrow 1$, when $a
\rightarrow 0$. In this way one can obtain, for example, new
solutions where  $\varphi$, $F$, $G$, $H$ depend not only on
$A=ia\partial_n$, but also on $\Delta=\partial_i\partial_i$.

Furthermore, two realizations with $\varphi_1(A)$ and
$\varphi_2(A)$, but with  $\psi_1(A)=\psi_2(A)=\psi(A)$, can be
connected by $S_{12}$:
$$
(\hat{x}_{\mu})_2=S_{12}(\hat{x}_{\mu})_1S_{12}^{-1},
$$
where
$$
S_{12}=exp\{x_i\partial_i(ln\varphi_2-ln\varphi_1)\}.
$$

%%%%%%%%%%%%%%%%%%%%%%%%%%%%%%%%%%%%%%%%%%%%%%%%%%%%%%%%%%%%%%
%%%%%%%%%%%%%%%%%%%%%%%%%%%%%%SECTION 7:
%%%%%%%%%%%%%%%%%%%%%%%%%%%%%%%%%%%%%%%%%%%%%%%%%%%%%%%%%%%%%%

\section {Hermiticity}

%%%%%%%%%%%%%%%%%%%%%%%%%%%%%%%%%%%%%%%%%%%%%%%%%%%%%%%%%%%%
%%%%%%%%%%%%%%%%%%%%%%%%%%%%%%%%%%%%%%%%%%%%%%%%%%%%%%
All relations of the type  \quad $[\hat{x},\hat{x}]$,\quad
$[M,M]$,\quad $[M,\hat{x}]$,\quad $[M,D]$,\quad $[D,D]$, \quad
$[D,\hat{x}]$,\quad eqs. (3), (7), (12), (21), (28),  are
invariant under the formal antilinear involution:
\begin{equation}
\hat{x}_\mu^{\dag}=\hat{x}_\mu,\quad D_\mu^{\dag}=-D_\mu , \quad
M_{\mu\nu}^{\dag}=-M_{\mu\nu},\quad c^{\dag}=\bar{c}, \quad
c\in{\bf C}.
\end{equation}
The order of elements in the product is inverted under the
involution. The commutative coordinates $x_\mu$ and their
derivatives $\partial_\mu$ also satisfy the involution property:
\quad $x_\mu^{\dag}=x_\mu,$\quad
$\partial_\mu^{\dag}=-\partial_\mu$. It is natural to ask whether
the realizations, eqs. (17)-(20), satisfy the involution property,
eq. (43). It is easy to verify that $\hat{x}_i^{\dag}=\hat{x}_i, $
and $D_\mu^{\dag}=-D_\mu$,\quad $M_{ij}^{\dag}=-M_{ij}$. However,
generally, $\hat{x}_n^{\dag}\neq \hat{x}_n,$\quad
$M_{in}^{\dag}\neq-M_{in}$.\\

We point out that all realizations can be made hermitian, i.e.
consistent with  eq. (43) by defining
\begin{equation}
\begin{array}{c}
\hat{x}_n^h=\frac{1}{2}(\hat{x}_n+\hat{x}_n^{\dag}),\\
\\
 M_{in}^{a.h.}=\frac{1}{2}(M_{in}-M_{in}^{\dag}).
\end{array}
\end{equation}
\\
All commutation relations are preserved by this redefinition of
$\hat{x}_n$ and  $M_{in}$ in any realization. Namely, if
$\hat{x}_\mu$, $M_{\mu\nu}$ is a realization of eqs. (3), (7),
(12), then $\hat{x}_\mu^{\dag}$, $-M_{\mu\nu}^{\dag}$ is also a
realization of the same relations. Moreover, any linear
combination  $\alpha\hat{x}_\mu+(1-\alpha)\hat{x}_\mu $\quad and
\quad $\alpha M_{\mu\nu}-(1-\alpha)M_{\mu\nu}^{\dag}$,
$\alpha\in\mathbf{R}$ satisfies the same relations,  eqs. (3),
(7), (12).  Specially, for $\alpha=\frac{1}{2},$ we obtain a
hermitian realization of NC space defined by  eqs. (3), (7), (12).\\
For example,
\\
\begin{equation}
\begin{array}{c}
\hat{x}_n^h=\frac{1}{2}(x_n\psi+iax_k\partial_k\gamma+\psi
x_n+ia\gamma\partial_kx_k)\\
\\
=x_n\psi+iax_k\partial_k\gamma+\frac{ia}{2}\psi'
+\frac{ia}{2}(n-1)\gamma.
\end{array}
\end{equation}
The simplest realization of $\kappa$-deformed space satisfying the
hermiticity property (43) is the left realization with
$\varphi=e^{-A}$, $\psi=1$ and $\gamma=0,$ i.e. (see sect. 8)
\begin{equation}
\begin{array}{c}
\hat{x}_i=x_ie^{-A}, \qquad \hat{x}_n=x_n,\\
\\
 D_i=\partial_i,\qquad
D_n=\frac{1}{a}sin(a\partial_n)+\frac{ia}{2}\Delta e^A,\\
\\
M_{ij}=x_i\partial_j-x_j\partial_i, \qquad
M_{in}=\frac{1}{a}x_isin(a\partial_n)-x_n\partial_ie^A+\frac{ia}{2}x_i\Delta
e^A, \\
\\
\Box=-\frac{2}{a^2}[cos(a\partial_n)-1]+\Delta e^A,
\end{array}
\end{equation}
with
\begin{equation}
e^{\pm ia\partial_n}f(x,...,x_n)=f(x,...,x_{n-1},x_n\pm ia).
\end{equation}
From the physical point of view, every $\varphi$-realization is
allowed and, in some sense, corresponds to choosing a "gauge" for
a  concrete calculation. Moreover, non-hermitian realizations (not
satisfying the hermiticity  properties, eq. (43)) are allowed for
concrete calculations [18]. For example, the symmetric realization
$\varphi_S(A)=\frac{A}{e^A-1}$ is not hermitian
$\hat{x}_n^+\neq\hat{x}_n$ [16-18], (see also sect. 8, eq. (67)).

%%%%%%%%%%%%%%%%%%%%%%%%%%%%%%%%%%%%%%%%%%%%%%%%%%%%%%%%%%%%%%
%%%%%%%%%%%%%%%%%%%%%%%%%%%%%SECTION 8:
%%%%%%%%%%%%%%%%%%%%%%%%%%%%%%%%%%%%%%%%%%%%%%%%%%%%%%%%%%%%%%%

\section {Realizations of star products}
%%%%%%%%%%%%%%%%%%%%%%%%%%%%%%%%%%%%%%%%%%%%%%%%%%%%%%%%%%%%%%%
%%%%%%%%%%%%%%%%%%%%%%%%%%%%%%%%%%%%%%%%%%%%%%%%%%%%%%%%%%%%%%%%

There exists an vector space isomorphism between (the coordinate
algebras of) the NC space $\mathbf{R}_{a}^n$ and of the Euclidean
space $\mathbf{R}^n$, depending on the function $\varphi(A)$
$(\psi=1$ or $\psi=1+2A)$ . In other words, for a given
realization described by $\varphi(A)$ there is a unique mapping
from the functions of the NC coordinates $\hat{x}_{\mu}$ to the
functions of the
commutative coordinates $x_{\mu}$.\\
Let us define the "vacuum" state:
\begin{equation}
\begin{array}{c}
|0>=1, \qquad D_{\mu}|0>=\partial_{\mu}|0>=0.
\end{array}
\end{equation}
Then we define a mapping from $f(\hat{x})$ to $f_\varphi(x)$ in a
given $\varphi$-realization as
\begin{equation}
f(\hat{x}_\varphi)|0>=f_\varphi(x).
\end{equation}
The functions $f(\hat{x})$ are defined as a formal power series in
NC coordinates. Note that all monomials in which $\hat{x}_{1}$
appear $m_{1}$ times, $\hat{x}_{2}$-$m_{2}$ times ,...,
$\hat{x}_{n}$-$m_{n}$ times, differ under permutations, i.e. there
are $m\choose m_{n}$ different monomials, where $m=\sum m_{\mu}$.
However, they are proportional to each other. A basis in the space
of monomials is fixed in a given $\varphi$-realization by
\begin{equation}\label{eq:Mbasis}
M_\varphi(\hat{x})|0>=M_\varphi(x)+P_\varphi(x),
\end{equation}
where $M_\varphi(\hat{x})$ is a linear combination of monomials of
the same type $(m_1,..., m_n)$  (i.e. $\hat{x}_1$ appearing $m_1$
times, $\hat{x}_2$-$m_2$ times, etc.), and $P_\varphi(x)$ is a
polynomial of lower order than $M_\varphi(x) $. We generally write
equation~(\ref{eq:Mbasis}) as
$$
\tilde{M}_\varphi(\hat{x})|0> =
[M_\varphi(\hat{x})+\tilde{P}_\varphi(\hat{x})]|0>= M_\varphi(x),
$$
where $\tilde{P}_\varphi(\hat{x})$ is a polynomial in $\hat{x}$ of
lower order than $M_\varphi(\hat{x})$. This means that a given
$\varphi$-realization induces a natural basis for monomials, i.e.
a natural ordering prescription, and vice versa, an ordering
uniquely defines the
$\varphi$-realization.\\
For example, equations (4),~(\ref{eq:Mbasis}) imply
 $\prod\hat{x}_i^{m_i}|0>=\prod x_i^{m_i}$ and for the mixed
second-order monomials we obtain
\begin{equation}
\begin{array}{c}
M_\varphi(\hat{x})=[1+\varphi'(0)]\hat{x}_{i}\hat{x}_{n}-\varphi'(0)\hat{x}_{n}\hat{x}_{i}
=\hat{x}_{i}\hat{x}_{n}-ia\varphi'(0)\hat{x}_{i},\\
\\
M_\varphi(\hat{x}_\varphi)|0>=x_i x_n.
\end{array}
\end{equation}
Let the $\tilde{M}_\varphi$ basis correspond to a given
$\varphi$-realization. Then
\begin{equation}
\begin{array}{c}
f(\hat{x})=f_\varphi(\hat{x})=\sum
c_{\varphi}\tilde{M}_\varphi(\hat{x}),\\
\\
f(\hat{x}_\varphi)|0>=f_\varphi(\hat{x}_\varphi)|0>=f_\varphi(x).
\end{array}
\end{equation}
Now we define a star product in a given $\varphi$-realization as
\begin{equation}
(f_\varphi\star_\varphi g_\varphi)(x) =
f_\varphi(\hat{x}_\varphi)g_\varphi(\hat{x}_\varphi)|0>=f_\varphi(\hat{x}_\varphi)g_\varphi(x).
\end{equation}
Generally,
\begin{equation}
\begin{array}{c}
x_{i}\star_\varphi
f(x)=(\hat{x}_\varphi)_{i}f(x)=x_{i}\varphi(A)f(x),\\
\\
x_{n}\star_\varphi
f(x)=(\hat{x}_\varphi)_{n}f(x)=[x_{n}\psi(A)+iax_{k}\partial_{k}\gamma(A)]f(x),
\end{array}
\end{equation}
and
\begin{equation}
\begin{array}{c}
f(x)\star_\varphi x_{i}=x_{i}\varphi(A)e^{A} f(x),\\
\\
f(x)\star_\varphi
x_{n}=[x_{n}\psi(A)+iax_{k}\partial_{k}(\gamma(A)-1)]f(x).
\end{array}
\end{equation}
\\
\\
%%%%%%%%%%%%%%%%%%%%%%%%%%%%%%%%%%%%%%%%%%%%%%%%%%%%%%%%
{\bf Realizations with $\psi=1$}\\
%%%%%%%%%%%%%%%%%%%%%%%%%%%%%%%%%%%%%%%%%%%%%%%%%%%%%%%%%%%%%
For $\psi=1$, $\gamma=\frac{\varphi'}{\varphi}+1$, we find a
closed form for the star product in the $\varphi$-realization
\begin{equation}
(f\star g)(z)=exp \Big\{ z_{i}\partial_{x_{i}}
[\frac{\varphi(A_{x}+A_{y})}{\varphi(A_{x})}-1]+z_{i}\partial_{y_{i}}
[\frac{\varphi(A_{x}+A_{y})}{\varphi(A_{y})}e^{A_{x}}-1]\Big\}f(x)g(y)\Big|_{x=z\atop
y=z},
\end{equation}
where $A_{x}=ia\frac{\partial}{\partial x_{n}}$ and
$A_{y}=ia\frac{\partial}{\partial y_{n}}.$ From eq. (56) it
follows generally
\begin{equation}
(g\star f)(z)\Big|_{\varphi(A)}=(f\star
g)(z)\Big|_{\varphi(A)e^A}.
\end{equation}
Using eq. (41), $\partial_i=D_i\varphi e^A$ and the expression for
coproduct $\triangle D_i$ (after eq.(28)), we can write  eq. (56)
for star product as
$$
(f\star g)
(z)=\Big(m\{exp[z_i(\triangle-\triangle_0)\partial_i](f\otimes
g)\}\Big)(x)\Big|_{x=z},
$$
where $\triangle_0\partial_i=\partial_i\otimes 1+1\otimes
\partial_i$ is the undeformed coproduct, and $m$ is the multiplication
map (product) in the Hopf algebra.\\
In the second order in $a$, from (56) we obtain
\begin{equation}
\begin{array}{c}
(f\star
g)(z)=f(z)g(z)+\Big\{z_{i}\Big[\Big(1+\varphi'(0)\Big)A_{x}
\partial_{y_{i}}+
\varphi'(0)\partial_{x_{i}}A_{y}\Big]\\
\\
+z_{i}\Big[\Big(
\frac{1}{2}+\varphi'(0)+\frac{1}{2}\varphi''(0)\Big)
A_{x}^2\partial_{y_{i}}\\
\\
 +\Big(\varphi''(0)-(\varphi'(0))^2\Big)(\partial_{x_{i}}A_{x}A_{y}+A_{x}A_{y}\partial_{y_{i}})
+\frac{1}{2}\varphi''(0)\partial_{x_{i}}A_{y}^2\Big]\\
\\
 +\frac{1}{2}z_{i}z_{j}\Big[\Big( 1+\varphi'(0)\Big)^2
A_{x}^2\partial_{y_{i}}\partial_{y_{j}} +2\varphi'(0)\Big(
1+\varphi'(0)\Big)\partial_{x_{i}}A_{x}A_{y}\partial_{y_{j}}\\
\\
+\Big(\varphi'(0)\Big)^2\partial_{x_{i}}\partial_{x_{j}}A_{y}^2\Big]\Big\}f(x)g(y)\Big|_{x=z\atop
y=z} \quad + \quad {\mathcal{O}}(a^3)
\end{array}
\end{equation}
and consequently\\
\begin{equation}
\begin{array}{c}
\Big(f\star g-g\star
f\Big)(x)=\Big(A_xf(x)\Big)\Big({\mathcal{N}}_{x}g(x)\Big)-
\Big({\mathcal{N}}_{x}f(x)\Big)\Big(A_{x}g(x)\Big)\\
\\
+\Big(\frac{1}{2}+\varphi'(0)\Big)\Big[\Big(A_{x}^2f(x)\Big)\Big({\mathcal{N}}_{x}^2g(x)\Big)-
\Big({\mathcal{N}}_{x}^2f(x)\Big)\Big(A_{x}^2g(x)\Big)\Big]\quad +
\quad {\mathcal{O}}(a^3),
\end{array}
\end{equation}
\\
where ${\mathcal{N}}_{x}=x_i\frac{\partial}{\partial x_i}$. We
point out that, generally, a factor
$A\otimes\mathcal{N}-\mathcal{N}\otimes A$
appears in all orders of the expansion $(f\star g-g\star f)(x)$.\\
 For a given $\varphi$-realization there is a monomial basis $M_{\varphi}$
satisfying eq. (50) with $P_\varphi(x)=0$, i.e.
$M_{\varphi}(\hat{x}_{\varphi})|0>=M_{\varphi}(x)$,
and vice versa. Let us consider three cases with $\psi=1$.\\
\\
%%%%%%%%%%%%%%%%%%%%%%%%%%%%%%%%%%%%%%%%%%%%%%%%%%%%%%%%
{\bf Left ordering}\\
%%%%%%%%%%%%%%%%%%%%%%%%%%%%%%%%%%%%%%%%%%%%%%%%%%%%%%%
\\
 If we define the $M_{\varphi}$ basis in such a
way that all $\hat{x}_{n}$ are at the most left in any monomial,
then it follows from eqs. (4), (50),
$\hat{x}_n^{m_n}\prod\hat{x}_i^{m_i}|0>=x_n^{m_n}\prod x_i^{m_i}$,
that  $\hat{x}_{n}=x_{n},$  i.e.
$\psi=1$,  $\gamma=0,$  $\varphi=e^{-A}$.\\
The star product for the left ordering is given by eq. (56) with
$\varphi_L=e^{-A}$, and alternatively by
\begin{equation}
(f\star_{\varphi_L}g)(x)=e^{-iax_i\partial_i^x\partial_n^y}
f(x)g(y)\Big|_{y=x}.
\end{equation}
%%%%%%%%%%%%%%%%%%%%%%%%%%%%%%%%%%%%%%%%%%
{\bf Right ordering}\\
%%%%%%%%%%%%%%%%%%%%%%%%%%%%%%%%%%%%%%%%%
\\
Similarly, the right ordering is defined so that $\hat{x}_{n}$ are
at the most right in any monomial. Then it follows from  eqs. (4),
(50), $(\prod\hat{x}_i^{m_i})\hat{x}_n^{m_n}|0>=(\prod
x_i^{m_i})x_n^{m_n}$, that  $ \hat{x}_{i}=x_{i}$,
$\psi=1$,  $\varphi=1$,  $\gamma=1$.\\
The star product for the right ordering is given by eq. (56) with
$\varphi_R=1$, and alternatively by
\begin{equation}
(f\star_{\varphi_R} g)(x)=e^{iay_i\partial_i^y\partial_n^x}
f(x)g(y)\Big|_{y=x}.
\end{equation}
\\
\\
%%%%%%%%%%%%%%%%%%%%%%%%%%%%%%%%%%%%%%%%%%%%%%%%%%%%%%%%%%%%%%%%
{\bf Symmetric ordering}\\
%%%%%%%%%%%%%%%%%%%%%%%%%%%%%%%%%%%%%%%%%%%%%%%%%%%%%%%%%%%%%%%%
\\
The general series expansion formula for the star product of the
Lie algebra type NC space, described by the structure constants
$C_{\mu\nu\lambda}$, was given in the symmetric ordering [31]. The
lowest-order symmetric monomials are
 $$
 \{\hat{x}_{i}\hat{x}_{j},\frac{1}{2}(\hat{x}_{i}\hat{x}_{n}+\hat{x}_{n}\hat{x}_{i}),
 \hat{x}_{n}^2\},
 $$
$$
\{\hat{x}_{i}\hat{x}_{j}\hat{x}_{k},\frac{1}{3}(\hat{x}_{i}\hat{x}_{j}\hat{x}_{n}+
\hat{x}_{i}\hat{x}_{n}\hat{x}_{j}+
\hat{x}_{n}\hat{x}_{i}\hat{x}_{j}),\frac{1}{3}(\hat{x}_{i}\hat{x}_{n}^2+
\hat{x}_{n}\hat{x}_{i}\hat{x}_{n}+\hat{x}_{n}^2\hat{x}_{i}),
\hat{x}_{n}^3\},\quad etc.
$$
To find a $\varphi$-realization corresponding to a symmetric
ordering, we impose the condition (50) as follows:
\begin{equation}
\sum_{\pi}\pi M(\hat{x})|0>={m\choose m_n}M(x)
\end{equation}
where the summation is over all different monomials $\pi
M(\hat{x})$ differing by permutations (summed with equal weights).
There are ${m\choose m_n}$ different monomials on the LHS, where
$m=\sum m_{\mu}$. Specially, for $m=k+1$, $m_n=k$, we have
\begin{equation}
\begin{array}{c}
\sum_{r=0}^k(\hat{x}_\varphi)_{n}^r(\hat{x}_\varphi)_{i}(\hat{x}_\varphi)_{n}^{k-r}|0>
= (k+1)x_{i}x_{n}^k ,\quad \forall k \in {\mathbf N}
\end{array}
\end{equation}
and use the relation (obtained by shifting $\hat{x}_i$ to the
left)
\begin{equation}
\begin{array}{c}
\sum_{r=0}^k\hat{x}_{n}^r\hat{x}_{i}\hat{x}_{n}^{k-r} =
\sum_{r=0}^k{k+1\choose
r+1}(ia)^r\hat{x}_{i}\hat{x}_{n}^{k-r},\\
\end{array}
\end{equation}
from which we find
\begin{equation}
\begin{array}{c}
\sum_{r=0}^k\sum_{p=0}^{k-r}{k+1\choose r+1}{k-r\choose
p}(ia)^{r+p}\varphi^{(p)}(0)x_{i}x_{n}^{k-r-p}=(k+1)x_{i}x_{n}^k,\\
\end{array}
\end{equation}
where $\varphi^{(p)}(0)$ is the $p$-th derivative calculated at $0$.\\
For \quad$l\ge1$, \quad we get
\begin{equation}
\sum_{p=0}^l{l+1\choose p}\varphi^{(p)}(0)=0.\\
\end{equation}
Solving the above recursive relations starting with
$\varphi(0)=1$, we obtain
\\
\begin{equation}
\begin{array}{c}
\varphi'(0)=-\frac{1}{2},\qquad\varphi''(0)=\frac{1}{12},\qquad\varphi'''(0)=0,
\quad etc.\\
\\
\varphi_s(A)=\sum_{p=0}^\infty\frac{\varphi_s^{(p)}(0)}{p!}A^{p}=\frac{A}{e^A-1}.
\end{array}
\end{equation}
\\
One can show that eq.~(62) is satisfied for every symmetrically
ordered monomial with the above $\varphi_S$,  eq.~(67).\\
 Note that
\begin{equation}
\begin{array}{c}
\gamma_s(A)=\frac{\varphi'_s(A)}{\varphi_s(A)}+1=-
\frac{\varphi_s(A)-1}{A},\\
\\
\varphi_s(A)e^A=\varphi_s(-A),\qquad \gamma_s(A)+\gamma_s(-A)=1,\\
\\
g(x)\star_{\varphi_S} f(x)\Big|_a=f(x)\star_{\varphi_S}
g(x)\Big|_{-a}.
\end{array}
\end{equation}
\\
Inserting $\varphi_L(A)=e^{-A},$  $\varphi_R(A)=1$,
 $\varphi_S(A)=\frac{A}{e^A-1}$ and the corresponding $\psi=1$,
$\gamma=\frac{\varphi'}{\varphi}+1$ functions for the left, right
and symmetric ordering, we obtain the  results for these three
special cases [18]. Generally, if $\psi=1$ then $\tilde
M_\varphi(\hat{x})=M_\varphi(\hat{x})$, i.e. $P_\varphi(x)=0$,
eqs. (4), (50).\\
\\
%%%%%%%%%%%%%%%%%%%%%%%%%%%%%%%%%%%%%%%%%%%%%%%
{\bf Realizations with $\psi=1+2A$}\\
%%%%%%%%%%%%%%%%%%%%%%%%%%%%%%%%%%%%%%%%%%%%%%%%%
\\
For the realizations  with $\psi=1+2A$, the condition described by
eq.~(50),
$$
\tilde M_\varphi(\hat{x})|0> = M_\varphi(x)
$$
can be fulfilled  generally if $\tilde M_\varphi(\hat{x})\neq
M_\varphi(\hat{x})$, i.e. if $P_\varphi(x)\neq 0$, eq.~(50).
Namely,
\begin{equation}
\hat{x}_n^k|0>=x_{n}[(1+2A)x_n]^{k-1}=x_n^k+P_{k-1}(x),\qquad
k\geq1,
\end{equation}
where $k\geq 1$, and $P_{k-1}(x)$ is a polynomial of order
$(k-1)$.\\
\\
 Generally, this holds for all realizations with $\psi\neq1$, including
the hermitian realizations with $\psi=1$, satisfying eq.~(43). For
example, we obtain (see eq.~(45))
\begin{equation}
\hat{x}_n^h|0>=x_n+\frac{ia}{2}\psi'(0)+\frac{ia}{2}(n-1)\gamma(0)\neq
x_n.
\end{equation}
The isomorphism $f$ to $f_\varphi$ is defined by
$f(\hat{x}_\varphi)|0>=f_\varphi(x),$ eqs.~(49), (52), and the
corresponding star product is defined by eq.~(53).

Our approach can be applied and extended to $\kappa$-deformed
Minkowski space. One can define  the Klein-Gordon and Dirac
equations for free fields and gauge theory in $\kappa$-deformed
space for an arbitrary $\varphi$-realization. There are still some
open problems concerning  the invariant integral and the
variational principle [32] that will be treated separately.

%%%%%%%%%%%%%%%%%%%%%%%%%%%%%%%%%%%%%%%%%%%%%%%%%%%%%%%%%%%%%%%%%%%%
%%%%%%%%%%%%%%%%%%%%%%%%%%%% SECTION 9:
%%%%%%%%%%%%%%%%%%%%%%%%%%%%%%%%%%%%%%%%%%%%%%%%%%%%%%%%%%%%%%%%%%%

\section {Conclusion}
We have presented a unified and simple method of constructing
realizations  of NC spaces in terms of commutative coordinates
$x_\mu$ and their derivatives $\partial_\mu$ of Euclidean space.
This method can also be applied  to spaces with arbitrary
signatures, especially to Minkowski-type spaces.

 Particularly, we have studied $\kappa$-deformed Euclidean space with
undeformed rotation algebra $SO_a(n)$. Dirac derivatives are
constructed  as a vector representation under $SO_a(n)$, and they
commute themselves $[D_\mu,D_\nu]=0$. Similarly, there is an
invariant operator $\Box$ such that $[M_{\mu\nu},\Box]=0$. We have
found two infinite new families of realizations described by
$\psi=1$ and $\psi=1+2A$ ($\psi=1$, $\varphi$ arbitrary and
$\psi=1+2A$, $\varphi=\frac{C-1}{C-\sqrt{\psi}}$, $C\neq 1$).\\
Furthermore, we have shown how these realizations can be extended
to satisfy the hermitian properties eq.~(43).

 We have constructed the star product for any realization. We point out that
all realizations are on an equal footing and any of the
realizations can be used for a concrete physical calculation, and
its meaning is similar to the case when a particular gauge is
chosen.

Finally, we have constructed $\kappa$-deformed  NC space of the
Lie algebra type, eq.~(31) with undeformed Poincar{\'e} algebra
(7), (21) and deformed  coalgebra (32), (33), in a unique way. Our
approach may be useful in quantum-gravity models, specially in 2+1
dimension. In this case, the corresponding Lie algebra is $SU(2)$
or $SU(1,1)$ [33-35]. For general Lie algebra, eq.~(1), there
exists a universal formula for mapping
$\hat{x}_{\mu}=x_{\alpha}\varphi_{\alpha\mu}(\partial_1,...,\partial_n),$
corresponding  to the totally symmetric ordering [36].

%%%%%%%%%%%%%%%%%%%%%%%%%%%%%%%%%%%%%%%%%%%%%%%%%%%%%%%%%%%%%%%%%%
{\bf Acknowledgments}
%%%%%%%%%%%%%%%%%%%%%%%%%%%%%%%%%%%%%%%%%%%%%%%%%%%%%%%%%%%%%%%%%
\\
One of us  (S.M.) thanks D. Svrtan and Z. {\v{S}}koda for useful
discussions. This work is supported by the Ministry of Science and
Technology of the Republic of Croatia.

%%%%%%%%%%%%%%%%%%%%%%%%%%%%%%%%%%%%%%%%%%%%%%%%%%%%%%%%%%%%%%%%%%%
%%%%%%%%%%%%%%%%%%%%%%%%%% REFERENCES
%%%%%%%%%%%%%%%%%%%%%%%%%%%%%%%%%%%%%%%%%%%%%%%%%%%%%%%%%%%%%%%%%
%%%%%%%%%%%%%%%%%%%%%%%%%%%%%%%%%%%%%%%%%%%%%%%%%%%%%%%%%%%%%%%%%%%

 \end{document}